\documentclass[preprint]{imsart}
\RequirePackage{amsthm,amsmath,amsfonts,amssymb}
\RequirePackage[authoryear]{natbib}
\RequirePackage[colorlinks,citecolor=blue,urlcolor=blue]{hyperref}
\RequirePackage{graphicx}

\startlocaldefs

\endlocaldefs

\begin{document}

\begin{frontmatter}
\title{On the Non-Gaussianity of Sea Surface Elevations}
\runtitle{On the Non-Gaussianity of Sea Surface Elevations}

\begin{aug}
\author[A]{\fnms{Alicia} \snm{Nieto-Reyes}\thanksref{t1}\ead[label=e1]{alicia.nieto@unican.es}}

\address[A]{Department of Mathematics, Statistics and Computer Science,\\
University of Cantabria, Spain,
\printead{e1}}

\thankstext{t1}{
A.N.-R.  was supported by grant MTM2017-86061-C2-2-P funded by  MCIN/AEI/ 10.13039/501100011033 and “ERDF A way of making Europe”.}


\end{aug}


\begin{abstract}
The sea surface elevations are generally stated as Gaussian processes in the literature. To show the inaccuracy of this statement, an empirical study of the buoys in the US coast at a random day is performed, which results  in rejecting the null hypothesis of Gaussianity in  over 80$\%$ of the cases. The analysis pursued relates to a recent one by the author in which the heights of sea waves are proved to be non-Gaussian. It is similar in that  the Gaussianity of the process is studied as a whole and not just of its one-dimensional marginal, as it is common in the literature. It differs, however, in that the analysis of the sea surface elevations is harder from a statistical point of view, as the one-dimensional marginals are commonly Gaussian, which is observed throughout the study. 
\end{abstract}

\end{frontmatter}

\section{Introduction}

Much attention in the literature is dedicated to the study of the sea surface height \citep{Forristall1978, Azais2005, Karmpadakis2020}, a function of the sea surface elevation which is generally obtained by making use of the zero-up or down crossing methodology. The sea surface height is relevant because of design and analysis of off-shore structures \citep{Haver1987} and  ships \citep{Mendes2021} and, therefore, the literature is large in terms of studying its distribution \cite{Tayfun,  Mori,  Stansell4, Stansell5, CasasPrat}. The sea surface height  has been modeled, for instance, as 
\begin{itemize}
\item a Rayleigh distribution \citep{Longuet1980, Jishad2021}, 
\item  a, more general, Weibull distribution \citep{Muraleedharan2007}, 
\item a  Forristall distribution \citep{Forristall1978}, 
\item a
{Naess distribution}  \citep{Naess},
\item a{ Boccotti distribution} \citep{Boccotti},
\item a{ Klopman distribution} \citep{Klopman}, 
\item a {van Vledder distribution} \citep{Vledder},
\item a {Battjes--Groenendijk distribution}  \citep{Battjes},
\item a { Mendez distribution}  \citep{Mendez},
{ or 
\item a LoWiSh II distribution}  \citep{Wu}.
\end{itemize}
 In \cite{Olas1} it is experimentally proved that the sea heights do not follow a Gaussian distribution. 
 
This study is dedicated, however, to the study of the sea surface elevation. The measurements of  sea surface elevation are obtained by buoys throughout the sea, which are later preprocessed to obtain the  sea heights. Sea surface elevations have been studied from a statistical point of view, studying its distribution \cite{srokosz1986joint}, the skewness of the distribution \cite{srokosz1986skewness},  and the modellization of the process \cite{hokimoto2014non, pena2018short}. Consideration has also being given to how to measure \cite{schulz2004measurement} and record the data \cite{collins2014recording}.
From an applied perspective, the literature contains works on sea surface elevations to, for instance, ship motion forecasting \cite{reichert2010x} and the development of sea surface elevation maps \cite{hessner2007sea}.

This work goes beyond the existing literature and it is dedicated to empirically prove that the distribution of the sea wave elevation is not necessarily Gaussian. 
From a statistical point of view, the importance of studying the sea surface elevation is high and lies in that it is a raw measurement.
While experimental studies show that the distribution of sea heights are clearly non-Gaussian, having a non-Gaussian one dimensional marginal, the non-Gaussianity of the sea surface elevation is not so obvious; which makes the problem more interesting. In fact, in proving the non-Gaussianity, it is here demonstrated that the cases that the literature considered  as Gaussian correspond to non-Gaussian processes with Gaussian one-dimensional marginals. To prove this, it is made here used of the random projection test \cite{nietoreyes2014}, a goodness of fit test that checks the Gaussianity of the process as a whole and not just of a finite order marginal, as other established test in the literature do; see, for instance \cite{epps1987,Lobato2004}.
The obtained findings are important due to the cases that the literature considered as Gaussian are the more numerous ones. These cases include very large waves and, in fact, according to \cite{Benetazzo2015}, very large waves might be much more frequent than commonly assumed.

The rest of the manuscript includes: The description of the studied dataset in Section \ref{Dataset} and of the applied methodology in Section \ref{metho}. The results of the analysis are described in Section \ref{results}. The analysis makes use of the nortsTest package of the R software.

\section{Datasets}\label{Dataset}


The Coastal Data Information Program ({https://cdip.ucsd.edu}) contains surface elevations measured by buoys that are along the cost of the US. For the present study, these measurement where downloaded on the
 24th of June 2021  from the web page  {https://thredds.cdip.ucsd.edu/thredds/catalog/cdip/ realtime/catalog.html}. In particular, the variable downloaded is that named  \emph{xyzZDisplacement}. The set of data used here  differs from that in \cite{Olas1} and it has not being used in the literature before.
 
 There are a total of 66 datasets, each corresponding to the collected time series of a station (buoy). Each buoy has an identification number, which is displayed in Table \ref{tab:tab1} (rows 1, 4, 7, 10, 13, 16, 19, 22, 25, 28 and 31)
There, it can also be observed the length of the associated time series, in the rows designated with the name  \emph{length}. The smallest length is depicted in bold, which is that of station 067. As it is just above a length of 30,000, each of the 66 datasets under study is restricted to a time series of length 30,000. The datasets consist of raw data, which contains unknown values. After taking out those, the length of the time series associated to each of the 66 buoys can also be observed from Table \ref{tab:tab1}. It is designated by the name \emph{studied}. Note that for buoys
 $$244, 197, 189 \mbox{ and } 092,$$ there is a line in the place designated for the variable \emph{studied}. This is because in those cases the whole 30,000
  first elements of the  time series have been unobserved. Consequently, those buoys are not here longer studied. It can also be observed from Table \ref{tab:tab1}   that there are seven buoys for which the whole 30,000
  first elements have been observed. One of those cases is that of buoy 249.
  
\begin{figure}
  \includegraphics[width=.495\linewidth]{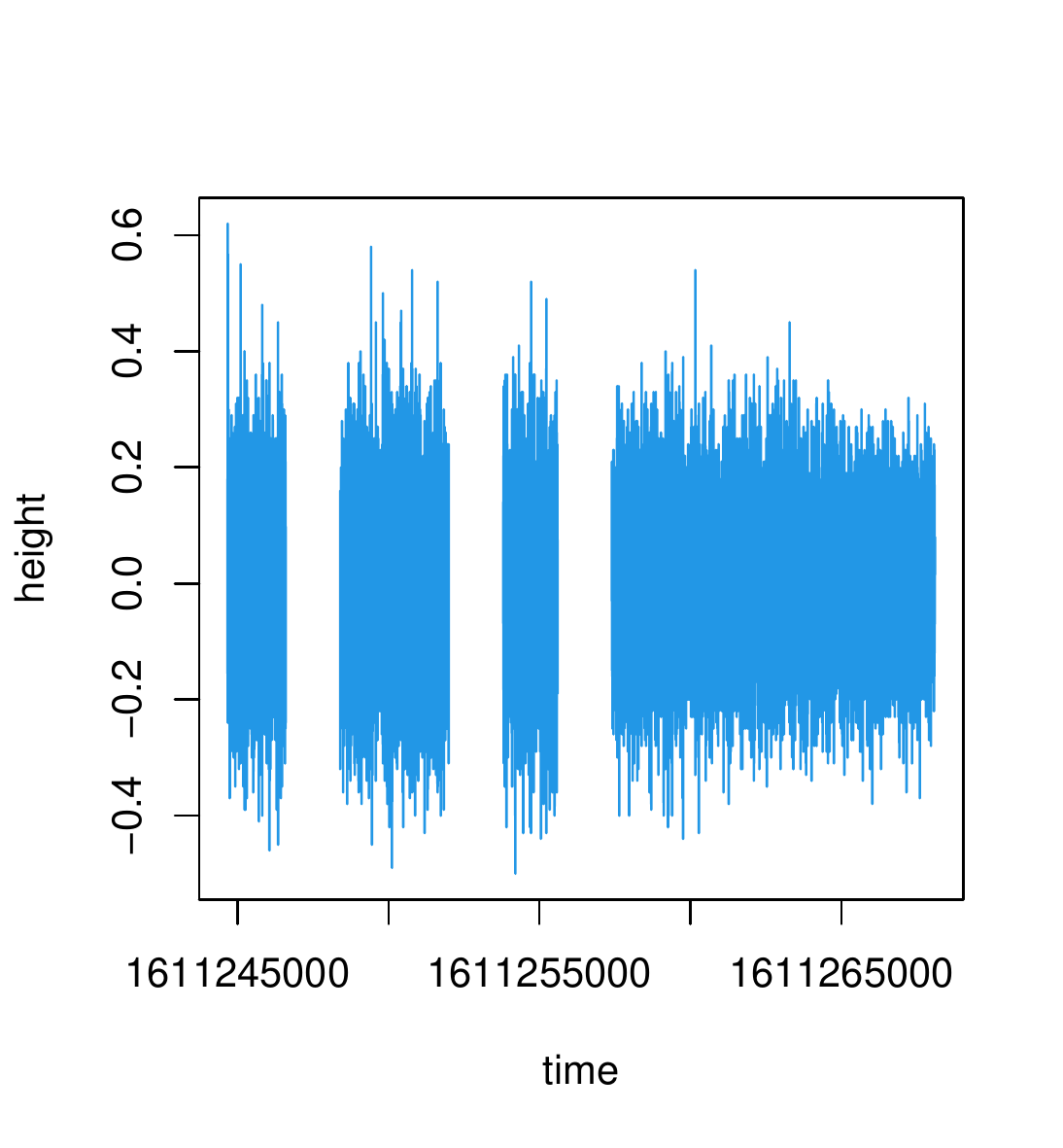} \includegraphics[width=.495\linewidth]{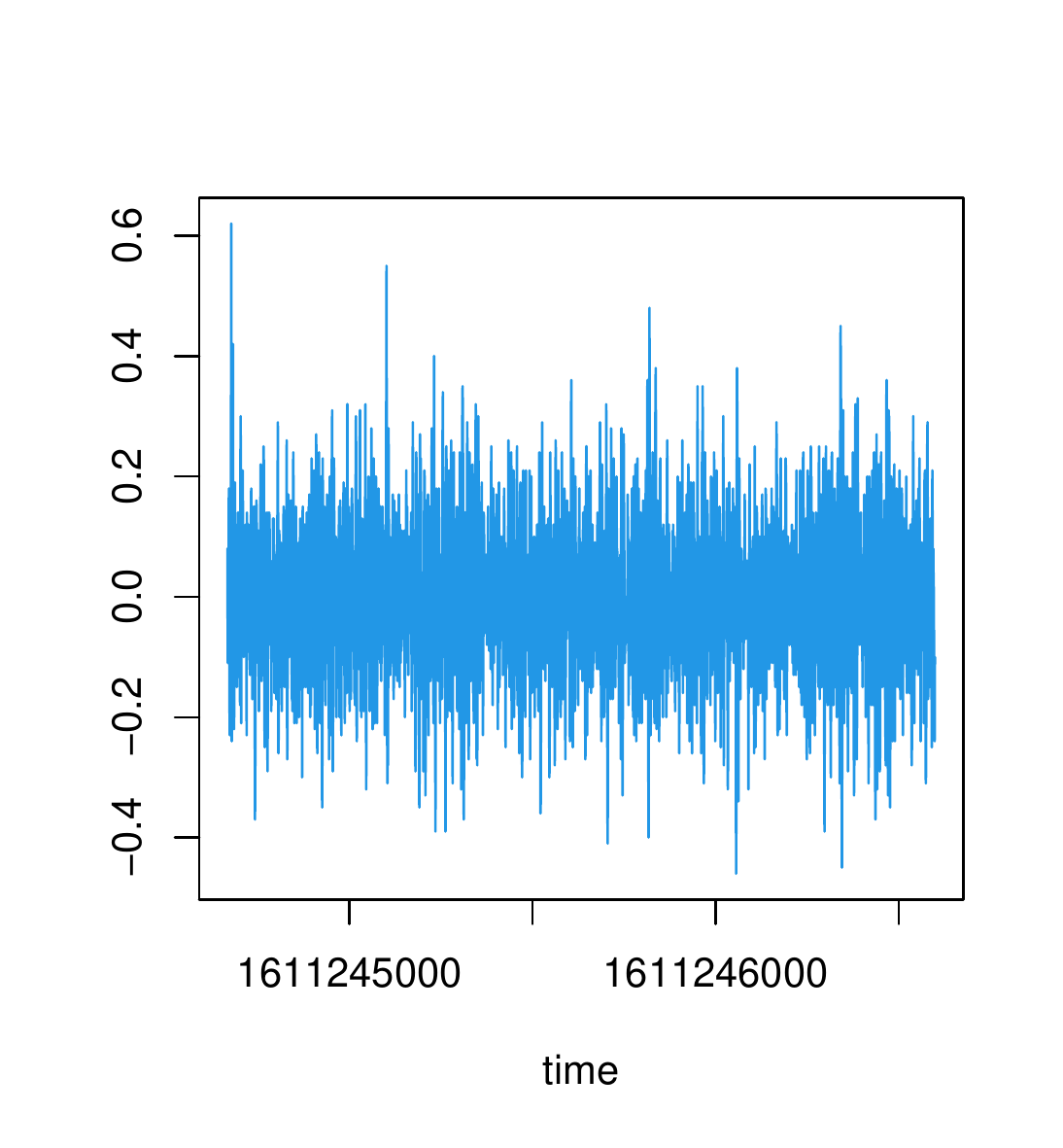}
  \caption{Left panel: representation of the studied time series for buoy 433. The three observed voids represent unobserved data. Right panel: representation of the first segment of the time series in the left panel. 
  Height in meters and time in seconds UTC.}\label{fig:fig1}
\end{figure}

\begin{table}[!ht]
\caption{The 66 available \emph{buoys} are labelled by an identification number in descending order. Below it is the  \emph{length} of the associated time series, the smallest being highlighted in bold. The  \emph{studied} label represents the length of the studied time series after selecting the first 30,000 time points  and eliminating the unobserved values. 
}
\begin{center}
  \setlength{\tabcolsep}{2.5mm}
 \begin{tabular}{lcccccc}
  \hline
 \bf buoy & 433 & 430 & 256 & 249 & 248 & 244 \\ 
 \bf length&  6177194 &67207849& 55277737& 31615487 &22526378 &81372840 \\
  \bf  studied & 23088&18480 & 20736& 30000&23856 &- \vspace{.1cm}\\ 
 \bf buoy & 243 & 240 & 239 & 238 & 236 & 233 \\
 \bf length& 29399210 &83692847& 28618154  &2511530& 13805738 &11474317\\
   \bf   studied &3184 &23088 &2304 &4608 & 7082&22272\vspace{.1cm} \\
 \bf buoy & 230 & 226 & 224 & 222 & 221 & 220 \\
 \bf length&    36864 &52940543 &47333545 &28634282 & 6659328 &35495593\\
 \bf       studied &27696 & 25392& 25392 & 30000& 27696&27696 \vspace{.1cm}\\
 \bf buoy & 217 & 215 & 214 & 213 & 209 & 203 \\
 \bf length& 20561066 &22436522 &79286411 &15161190 &18012842 &10925950\\
  \bf        studied & 20614& 30000& 27696&27696 &30000 &27696 \vspace{.1cm}\\
 \bf buoy & 201 & 200 & 198 & 197 & 196 & 192 \\
 \bf length& 40785577 & 7713962& 25192106&  2520576 &80495016& 33623807\\
  \bf          studied & 30000& 23855& 16006&- &11520 & 30000\vspace{.1cm}\\
 \bf buoy & 191 & 189 & 187 & 185 & 181 & 179 \\
 \bf length& 31265450& 35170729 &39129001  &8953514 &80459761& 58130089\\
  \bf            studied &27696 &- & 11690&27696 &27696 & 25222\vspace{.1cm}\\
 \bf buoy & 171 & 168 & 162 & 160 & 158 & 157 \\ 
 \bf length& 42015913 &14455296 &41025193 &  854954& 97940734 &53277865\\
 \bf               studied &23856 & 27696& 20784&27696 & 27696& 20614\vspace{.1cm}\\
 \bf buoy &155 & 154 & 153 & 150 & 147 & 144 \\
 \bf length&  4737194& 32389802 & 7939754& 62576809 &55812265 &18123434\\
 \bf                 studied & 18480& 27696& 19968& 30000 & 23088&25514 \vspace{.1cm}\\
 \bf buoy &143 & 142 & 139 & 134 & 132 & 121 \\
 \bf length& 27956906 &16902314& 43386793& 37933225 &39679657 & 859392\\
   \bf                 studied &30000 &3072 & 20784&25392 & 13104&23088 \vspace{.1cm}\\
 \bf buoy & 106 & 100 & 098 & 094 & 092 & 076 \\
 \bf length&    65280& 30917546 & 7206912& 12358826 &40140457 &44669119\\
   \bf                   studied &9613 &27696 &7082 &23088 & -& 27696\vspace{.1cm}\\
 \bf buoy & 071 & 067 & 045 & 036 & 029 & 028 \\
 \bf length& 78400512 &   {\bf 32256}  & 990890  & 141312 & 9384362 &17022122\\
     \bf                   studied &23088 &18480 &25392 & 30000& 30000& 25344\\
\hline
\end{tabular}
\end{center}
\label{tab:tab1} 
\end{table}

In the left panel of Figure \ref{fig:fig1}, it is displayed the studied data for buoy 433. As explained when commenting on  Table \ref{tab:tab1}, this dataset results from restricting the 6,177,194 observations stored for buoy 433 and taking the ones corresponding to the first 30,000 time points. As it is obvious from the plot, the first 30,000 time points contain unobserved elements. Indeed, only 23,088 observations have been made (see Table \ref{tab:tab1}). From the left panel of the figure, it is also observable that the unobserved data splits the time series in four parts.  The right panel of Figure \ref{fig:fig1} is a zoom of the left panel containing solely the part to the left of the time serie.  Coordinated universal times (UTC)  in seconds are used for the two panels in the figure.
The studied measurements of buoy 433 started being measured at time 1,611,244,667 (see Table \ref{tab:tab2}), which is  $$\mbox{ Thursday 21st of January of  2021 at  the 15:00 hours 57 minutes and 47 seconds }$$ in Greenwich mean time (GMT).  The last studied measurement of that buoy was recorded at 1,611,268,104, which is later the same day, at the $$\mbox{ 22:00 hours 28 minutes and 24 seconds}$$ in GMT. In the x-axis of the plots in the figure, it appears four  time points that have been translated in Table \ref{time}.

 \begin{table}[!ht]
\caption{For each of the 62 studied buoys, display of the starting and ending time points in seconds UTC of the recordings. The smallest and largest time points are highlighted in bold.}
 \begin{center}
  \setlength{\tabcolsep}{2.2mm}
\begin{tabular}{cccccc}
  \hline
 \bf  buoy &  \bf start time&   \bf end time & \hspace{.5cm}   \bf buoy &  \bf start time&  \bf end time\\
  &seconds UTC&seconds UTC& \hspace{.5cm} &seconds UTC&seconds UTC\\
   \hline \\
  433 & 1611244667  & 1611268104  &  \hspace{.5cm} 181 & 1597067867  & 1561681704 \\
  430 & 1617209867  & 1617233304  & \hspace{.5cm}  179 & 1600117067  & 1622597437\\
  256 & 1624062470  & 1624074188  &  \hspace{.5cm} 171 & 1606739270  & 1604873585
\\
    249 & 1623769067  & 1623792504  &  \hspace{.5cm} 168 & 1561658267  & 1618540104
\\
     248 & 1616997470  & 1617009188  &  \hspace{.5cm} 162 & 1622574000  & 1592702904
\\
     243 & 1589644667  & 1593206904  & \hspace{.5cm}  160 & 1604861867 &  1616030738
\\
    240 & 1593183467  & 1614911304  &  \hspace{.5cm} 158 & 1618516667  & 1610494104
\\
     239 & 1614887867  & 1606923037  & \hspace{.5cm}  157 & 1592679467  & 1609976867
\\
     238 & 1606899600  & 1571545704  &  \hspace{.5cm} 155 & 1616007301  & 1562624904
\\
    236 & 1571522267  & 1612387707  &  \hspace{.5cm} 154 & 1610470667  & 1607038104
\\
    233 & 1612364270  & 1623883518  &  \hspace{.5cm} 153 & 1609953430  & 1608491585
\\
    230 & 1623871800  & 1593566904  &  \hspace{.5cm} 150 & 1562601467  & 1596835704
\\
    226 & 1593543467  & 1594931304  &  \hspace{.5cm}  147 & 1607014667  & 1620458907
\\
    224 & 1594907867  & 1589239704  &  \hspace{.5cm}  144 & 1608479867  & 1602196104
\\
    222 & 1589216267  & 1611361704 &    \hspace{.5cm} 143 & 1596812267  & 1587587304
\\
    221 & 1611338267  & 1602725304  &   \hspace{.5cm} 142 & 1620435470  & 1600216988
\\
    220 & 1602701867  & 1610407704  &   \hspace{.5cm} 139 & 1602172667  & {\bf 1624302037}
\\
    217 & 1610384267  & 1580963304  &   \hspace{.5cm} 134 & 1587563867  & 1615602504
\\
    215 & 1580939867  & 1575678504  &   \hspace{.5cm} 132 & 1600205270  & 1613769185
\\
    214 & 1575655067  & 1618363704  &   \hspace{.5cm} 121 & 1624278600  & 1622604504
\\
   213 & 1618340267  & 1599262104  &  \hspace{.5cm}  106 & 1615579067  & 1602210504
\\
    209 & 1599238667  & 1620865704  &  \hspace{.5cm}  100 & 1613757467 &  1601598504
\\
    203 & 1620842267  & 1582943304 &    \hspace{.5cm} 098 & 1622581067  & 1560994104
\\
    201 & 1582919867  & 1581073107 &    \hspace{.5cm} 094 & 1602187067 &  1606966104
\\
    200 & 1581049670  & 1623885185 &   \hspace{.5cm}  076 & 1601575067  & 1581380904
\\
   198 & 1623873467  & 1608373837  &   \hspace{.5cm} 071 & {\bf 1560970667}  & 1572060504
\\
    196 & 1608350400 &  1579152504  &  \hspace{.5cm}  067 & 1606942667  & 1619742504
\\
    192 & 1591718267  & 1610760504  &   \hspace{.5cm} 045 & 1599843600  & 1611268104
\\
    191 & 1579129067  & 1587371304  &   \hspace{.5cm} 036 & 1581357467  & 1617233304
\\
    187 & 1610737067  & 1600140504  &  \hspace{.5cm}  029 & 1572037067 &  1624085907
\\
    185 & 1587347867  & 1606762707  &  \hspace{.5cm}  028 & 1619719067  & 1623792504 \\
\hline
\end{tabular}
\end{center}
\label{tab:tab2} 
\end{table}

 \begin{table}[!ht]
 \caption{Translation of UTC time points in seconds to regular nomenclature in GMT.}
 \begin{center}
 \setlength{\tabcolsep}{12.2mm}
\begin{tabular}{cc}
  \hline
  \bf seconds UTC & \bf GMT \\
     \hline \\
  1611245000& Thursday January 21st 2021 $16:03:20$ \\
  1611246000& Thursday January 21st 2021 $16:20:00$ \\
  1611255000& Thursday January 21st  2021 $18:50:00$ \\
  1611265000& Thursday January 21st 2021 $21:36:40$ \\
  \hline
\end{tabular}
 \label{time} 
   \end{center}
\end{table}

Apart from the ones of station 433, Table \ref{tab:tab2} displays the starting and ending recording times for each of the studied stations. The largest time value is 1,624,302,037, representing the $$\mbox{ Monday 21st of June  2021 at the 19:00 hours and 37 seconds GMT.}$$ This is highlighted in bold in the table. This is the end time for the recording of buoy 139, whose starting time is 1,602,172,667, i.e., the $$\mbox{ Thursday 8th of October 2020 at the 15:00 hours 57 minutes and 47 seconds GMT.}$$
Meanwhile, the smallest starting time point is 1,560,970,667, which represents the $$\mbox{ Wednesday 19th of June 2019 at the 18:00 hours 57 minutes and 47seconds GMT;}$$ and which is also highlighted  in bold in the table. It corresponds to the buoy with identification number 071, whose end time point is 1,572,060,504, i.e. the $$\mbox{ Saturday 26th of October 2019 at the 03:00 hours 28 minutes and 24 seconds GMT.}$$

\section{Methodology}\label{metho}

Given $X_t$  a real valued random variable for each $t\in \mathbb{Z},$ $$X:=\{X_t\}_{t\in \mathbb{Z}}$$ is a stochastic process \cite{Coleman1974}. Most common hypotheses on stochastic processes are those of stationarity \cite{zbMATH03244325} and Gaussianity \cite{KOZACHENKO201671}. $X$ is stationary if 
\begin{itemize}
\item $\mbox{E}[X_t]=\mbox{E}[X_{t+k}]$ for all $k, {t}\in  \mathbb{Z},$ where $\mbox{E}$ denotes the expectation function,
 \item $\mbox{Cov}(X_t,X_k)=\mbox{Cov}(X_{t-k},X_0)$ for all $k, {t}\in  \mathbb{Z},$ where $\mbox{Cov}$ denotes the covariance function and
 \item $\mbox{Var}[X_t]<\infty$ for all ${t}\in  \mathbb{Z},$ where $\mbox{Var}$ denotes the variance. 
 \end{itemize}
$X$ is Gaussian if 
$$
(X_{t_1}, \ldots, X_{t_n}) \mbox{ is a Gaussian random vector for all  } n\in \mathbb{N}.
$$
 It occurs that a stationary Gaussian process is strictly stationarity. $X$ is strictly stationary if $$(X_{t_1}, \ldots, X_{t_n}) \mbox{ and } (X_{t_{1+k}}, \ldots, X_{t_{n+k}})$$ are equally distributed for all $n\in \mathbb{N}$ and $k, {t_1}, \ldots, {t_n}\in  \mathbb{Z}.$ Consequently, given a stationary process $X,$  it is Gaussian if 
\begin{equation}\label{G}
(X_{t}, \ldots, X_{t}) \mbox{ is a Gaussian random vector for all  } t\in \mathbb{N}.
\end{equation}

\subsection{Tests for stationarity}\label{stat}
This manuscript is about testing the Guassianity of stocastic processes. Typically, those tests assume that the process is stationary. Thus, this assumption has to be previously checked. For that, the most common tests in the literature are 
\begin{enumerate}
 \item Ljung-Box test  \citep{Box},
\item
 Augmented Dickey-Fuller test \citep{dickey1984},  
 \item Phillips-Perron  test  \citep{Perron1988} and
 \item  kpps  test  \citep{KppsI1992} .
 \end{enumerate}
 For the first three tests, the tests can be simplified as contrasting the null hypothesis
 \begin{equation}\label{01}H_{0,1}: X \mbox{ is non stationary}\end{equation}
against the alternative
$$H_{a,1}: X \mbox{ is  stationary}$$
while the kpps  test results in  the null hypothesis 
 \begin{equation}\label{02}H_{0,2}: X \mbox{ is stationary}\end{equation}
against the alternative
$$H_{a,2}: X \mbox{ is  non stationary}.$$

The hypotheses are tested in different ways. For instance, Ljung-Box test makes use of the autocorrelation function, which, at lag $k$ for a stationary process is $$\frac{\mbox{Cov}(X_t,X_{t+k})}{\mbox{Var}(X_t)}.$$ This is observable from its statistic: $$n(n+2)\sum_{k=1}^{h}\frac{\hat\rho^2_k}{n-k},$$ where  $\hat\rho_k$ denotes the sample autocorrelation at lag $k$ and $n$ the sample size. Note that it depends on a constant $h.$

\subsection{Tests for Gaussianity}\label{gau}
Most tests for Gaussianity of stochastic processes assume the process is stationary and test whether a finite marginal distribution of the process is Gaussian, generally, the one-dimensional marginal. 
That is, instead of testing whether \eqref{G} is satisfied, these tests contrast the null hypothesis
 \begin{equation}\label{03}H_{0,3}: X_{t} \mbox{ is a Gaussian random variable}\end{equation}
 against the alternative  $$H_{a,3}: X_{t} \mbox{ is not a Gaussian random variable}$$
 by checking whether $X_t$ is a Gaussian random variable.
  Let us reflect  that, because of the stationarity, the distribution of  $X_{t}$ is the same for all $t\in  \mathbb{Z};$ that is, it is independent of $t.$
 
 Common tests to check the Gaussianity of a real valued random variable require a sample of independent and identically distributed random variables \cite{Dagostino1987}. As this work deals with stochastic processes, the independence assumption is not verified. However, there are also many tests for this situation. Here, it is  made use of the Epps test \cite{epps1987}, which checks that the characteristic function of the one-dimensional distribution of the process is that of a Gaussian distribution, and of the Lobato and Velasco test \cite{Lobato2004}, which checks that the third and fourth order moments of the one-dimensional distribution of the process are those of a Gaussian distribution. 
 
 If the null hypothesis $H_{0,3}$ is rejected, with the above mentioned tests, the null hypothesis
  \begin{equation}\label{04}H_{0,4}: X \mbox{ is a Gaussian process}\end{equation}
  is rejected against the alternative  $$H_{a,4}: X \mbox{ is not a Gaussian process}.$$ However, it may occur that $$X_{t} \mbox{ is a Gaussian random variable}$$ while $$X \mbox{ is not a Gaussian process}.$$ The above mentioned tests are at nominal level again this type of alternatives. For this, it is used here the random projection test \cite{nietoreyes2014}, which test the Gaussianity of the whole distribution of the process and not just of a finite dimensional marginal.  For  elaboration on it, see  Subsection \ref{RP}.

\subsubsection{Random projection test}\label{RP}

The random projection test was introduced in  \cite{nietoreyes2014} as a tool to test the Gaussianity of stationary processes that is able to reject the null hypothesis of Gaussianity \eqref{04} against alternatives with Gaussian finite-dimensional marginals. The procedure is based on a result in \cite{Cuesta2007} that implies that if $$\langle \{X_j\}_{j\leq t} , d\rangle,$$ with $d$ drawn from a Dirichlet distribution \cite{Pitman}, is Gaussian, then $$\{X_j\}_{j\leq t}$$ is Gaussian. Note that due to the stationarity assumption, the Gaussianity of  $\{X_j\}_{j\leq t}$ is equivalent to \eqref{G}. In what follows,   the procedure is explained in detail.

Let $$\lambda_1, \lambda_2>0$$ be two parameters. Making use of the following stick-breaking procedure, a Dirichlet distribution  is considered: 
\begin{enumerate}
\item Let $$\beta(\lambda_1, \lambda_2)$$ denote a beta distribution with parameters $\lambda_1, \lambda_2.$
\item Let $d_0$ be drawn from  the distribution $\beta(\lambda_1, \lambda_2).$ Note that $$d_0\in[0,1].$$
\item For any $k\in\mathbb{N},$ the natural numbers, let $d_k$ be the result of multiplying 
$$1-\sum_{i=0}^{k-1}d_i$$ and an element drawn independently from he distribution $\beta(\lambda_1, \lambda_2).$  Note that $$d_k\in[0,1-\sum_{i=0}^{k-1}d_i].$$
\end{enumerate}

Let $X$ be a stationary  process. The associated projected process based on $\{d_k\}_{k\in\mathbb{N}}$ is $$Y:=\{Y_t\}_{t\in \mathbb{Z}}$$ with
$$Y_t:=\sum_{i=0}^{\infty}d_iX_{t-i}.$$
Then, making use of this randomly projected process, it suffices to apply to it a test for the null hypothesis of Gaussianity \eqref{03}. 

The selection of the parameters $\lambda_1, \lambda_2$ is important. It is explained in  \cite{nietoreyes2014} that values such us $$\lambda_1=100 \mbox{ and } \lambda_2=1$$ result in an projected process $Y$ similar to $X.$ However, values such us $$\lambda_1=2 \mbox{ and } \lambda_2=7$$ result in projected processes different from $X$ while providing an effective method.

\subsection{False discovery rate}\label{FDR}
When multiple tests are perform, the multiplicity has to be taken into account. For that it is used here the false discovery rate \citep{Benjamin2001}.  The false discovery rate aims at controling the expected proportion of falsely rejected hypothesis. It was first introduced in \cite{Benjamin1995} to take into account the multiplicity of independent tests. In \citep{Benjamin2001} it was established that the definition in \citep{Benjamin1995} remains valid for certain types of dependency. However, for general dependent cases  \citep{Benjamin2001} has to be applied.


\section{Results of the analysis}\label{results}
This section analyzes whether each of  the 62 datasets provided in Section \ref{Dataset}, one per buoy, is drawn from a Gaussian process. For that,  the tests described in Section \ref{gau} are used here. As they require the stationarity assumption for the process, making use of the tests provided in \ref{stat}, it is first checked whether  each of  the datasets is drawn from a stationary process. 

The results obtained from checking the stationarity are displayed in Table \ref{st}. Only one result is provided by test because  the same one is obtained for each of the 62 datasets. Thus, p-values smaller than .01 are obtained for the {Augmented Dickey-Fuller} test and the {Phillips-Perron} test. P-values that are approximately zero are obtained for the {Ljung-Box} test and p-values larger than .1 for the {kpps} test.
Note that in the first three tests the null hypothesis  of non-stationarity is tested, as in \eqref{01}, and in the fourth it is the null hypothesis  of stationarity, as in \eqref{02}. Thus, it can be assumed that the studied datasets are drawn from stationary processes, and check their Gaussianity under the mentioned assumption.

 \begin{table}[!htbp]
 \caption{Obtained p-values for each of the 62 studied datasets under four different stationarity tests:  \textit{Augmented Dickey-Fuller} (first column), \textit{Phillips-Perron} (second column), \textit{Ljung-Box test}  (third column) and   \textit{kpps} (fourth column).  The null hypothesis is of stationarity for the  \textit{kpps} test and of non-stationarity for the other three.
 }
 \begin{center}
   \setlength{\tabcolsep}{2.2mm}
\begin{tabular}{ccccc}
  \hline
  && \bf Tests&&\\
& \bf \textit{Augmented Dickey-Fuller}&  \bf  \textit{Phillips-Perron}&  \bf \textit{Ljung-Box}  &  \bf  \textit{kpps}\\ \hline \\
 \bf p-\mbox{value}&$< .01$   & $<.01$  & 0 &$ >.1$
   \\
\hline
\end{tabular}
\end{center}
 \label{st} 
\end{table}

In order to study the Gaussianity of the datasets under study, it is first analyzed the one-dimensional marginal distribution of the process. This is because a rejection of the null hypothesis \eqref{03} implies the sought rejection of the whole distribution of the process, in \eqref{04}.
For  analyzing the one-dimensional marginal distribution, it is made use of the Epps and Lobato and Velasco tests commented in Section \ref{metho}. The results are displayed in Table \ref{tab:tab3}. There, for each dataset, associated to a buoy (columns 1 and 5), it can be observed the p-values resulting from applying the Epps test (columns 2 and 6) and the Lobato and Velasco test (columns 3 and 7). As multiplicity has to be taken into account, columns 4 and 8 display the FDR values. It can be observed from the table that half of the 62 FDR values are smaller than 0.05 studied. They have been highlighted in   bold. 
If  the less conservative FDR introduced in \citep{Benjamin1995} had been used, the number of rejections would have increased.  For instance, the null hypothesis \eqref{03} for buoys 185 and 028 would also have  been rejected.
If multiplicity had not been taken into account at all and  the null hypothesis \eqref{03} were rejected when  the minimum of the two p-values was smaller than 0.05, the number of rejections would have increased to 40.

 \begin{table}[t]
\caption{P-values resulting from applying the Epps test (columns 2 and 6)  and  the Lobato and Velasco test (columns 3 and 7) per dataset associated to each buoy (columns 1 and 5). FDR (columns 4 and 8) combination, for dependent p-values, of the two p-values per buoy, with the ones smaller than 0.05 highlighted in bold.
}
 \begin{center}
   \setlength{\tabcolsep}{3.4mm}
\begin{tabular}{rrrrrrrr}
  \hline
 \bf buoy& \bf  Epps &  \bf L.-V.& \bf FDR    \hspace{.25cm}  &   \hspace{.25cm}  \bf buoy&  \bf Epps &  \bf L.-V.& \bf FDR  \\ \hline \\
433  & .000  & .000  &\bf  .000  \hspace{.25cm}  &   \hspace{.25cm} 181  & .000  & .000  & \bf .000
\\
    430  & .027  & .000  & \bf  .000  \hspace{.25cm}  &   \hspace{.25cm}  179  & .000  & .000  & \bf  .000
\\
    256  & .004  & .000  &\bf   .000   \hspace{.25cm}  &   \hspace{.25cm}  171  & .000  & .000  &\bf   .000
\\
    249  & .000  & .000  &\bf   .000   \hspace{.25cm}  &   \hspace{.25cm} 168  & .001  & .000  &\bf   .000
\\
    248  & .524  & .044  &   .132   \hspace{.25cm}  &   \hspace{.25cm} 162  & .331  & .046  &  .138
\\
    243  & .063  & .044  &   .095   \hspace{.25cm}  &   \hspace{.25cm} 160  & .111  & .011  &\bf   .033
\\
    240  & .017  & .000  & \bf  .000   \hspace{.25cm}  &   \hspace{.25cm} 158  & .290  & .178  & .436
\\
    239  & .297  & .780  & .891   \hspace{.25cm}  &   \hspace{.25cm} 157  & .116  & .599  & .349
\\
    238  & .000  & .552  & 1.656   \hspace{.25cm}  &   \hspace{.25cm} 155  & .066  & .063  &   .099
\\
   236  & .708  & .142  & .426   \hspace{.25cm}  &   \hspace{.25cm} 154  & .011  & .000  &\bf   .000
\\
   233  & .004  & .000  &\bf   .000   \hspace{.25cm}  &   \hspace{.25cm} 153  & .028  & .000  &\bf   .000
\\
   230  & .000  & .000  &\bf   .000   \hspace{.25cm}  &   \hspace{.25cm} 150  & .013  & .000  & \bf  .000
\\
   226  & .254  & .030  &   .090   \hspace{.25cm}  &   \hspace{.25cm} 147  & .055  & .001  &\bf   .003
\\
   224  & .000  & .000  & \bf  .000   \hspace{.25cm}  &   \hspace{.25cm} 144  & .000  & .000  &\bf   .000
\\
   222  & .821  & .788  & 1.231   \hspace{.25cm}  &   \hspace{.25cm} 143  & .261  & .005  &\bf   .015
\\
   221  & .000  & .000  &\bf   .000   \hspace{.25cm}  &   \hspace{.25cm} 142  & .586  & .093  & .279
\\
   220  & .739  & .356  & 1.068   \hspace{.25cm}  &   \hspace{.25cm} 139  & .292  & .646  & .877
\\
   217  & .000  & .000  &\bf   .000   \hspace{.25cm}  &   \hspace{.25cm} 134  & .188  & .000  & \bf  .000
\\
   215  & .057  & .035  &   .086   \hspace{.25cm}  &   \hspace{.25cm} 132  & .017  & .000  & \bf  .000
\\
   214  & .068  & .004  &\bf   .012   \hspace{.25cm}  &   \hspace{.25cm} 121  & .001  & .000  & \bf  .000
\\
   213  & .002  & .001  & \bf  .003  \hspace{.25cm}  &   \hspace{.25cm}  106  & .574  & .887  & 1.4330
\\
   209  & .817  & .000  & \bf  .000  \hspace{.25cm}  &   \hspace{.25cm}  100  & .122  & .114  & .183
\\
   203  & .948  & .857  & 1.422  \hspace{.25cm}  &   \hspace{.25cm}  098  & .280  & .914  & .840
\\
   201  & .329  & .227  & .493  \hspace{.25cm}  &   \hspace{.25cm}  094  & .521  & .056  & .168
\\
   200  & .000  & .000  &\bf   .000   \hspace{.25cm}  &   \hspace{.25cm} 076  & .068  & .110  & .165
\\
   198  & .090  & .000  & \bf  .000   \hspace{.25cm}  &   \hspace{.25cm} 071  & .261  & .020  &   .060
\\
   196  & .235  & .062  & .186  \hspace{.25cm}  &   \hspace{.25cm}  067  & .790  & .426  & 1.184
\\
   192  & .093  & .000  &\bf  .000  \hspace{.25cm}  &   \hspace{.25cm} 045  & .587  & .222  & .666
\\
   191  & .535  & .079  &  .237    \hspace{.25cm}  &   \hspace{.25cm} 036  & .002  & .011  &\bf   .007
\\
   187  & .803  & .334  &1.002   \hspace{.25cm}  &   \hspace{.25cm} 029  & .472  & .990  & 1.416
\\
   185  & .020  & .287  &   .059   \hspace{.25cm}  &   \hspace{.25cm} 028  & .359  & .017  &  .051
   \\
\hline
\end{tabular}
\end{center}
\label{tab:tab3} 
\end{table}

\begin{figure}[t]
\begin{center}
  \includegraphics[width=.495\linewidth]{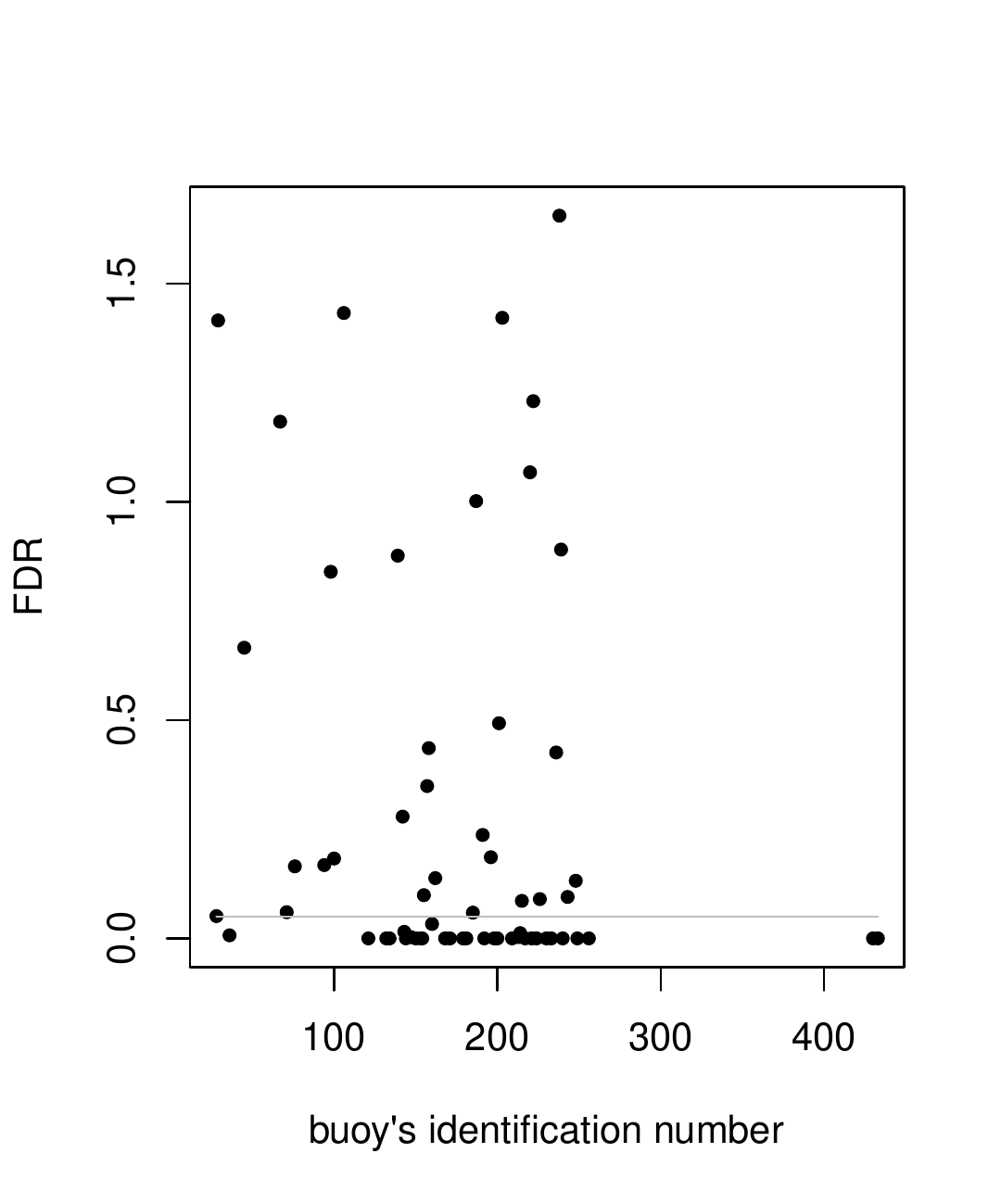} 
  \includegraphics[width=.495\linewidth]{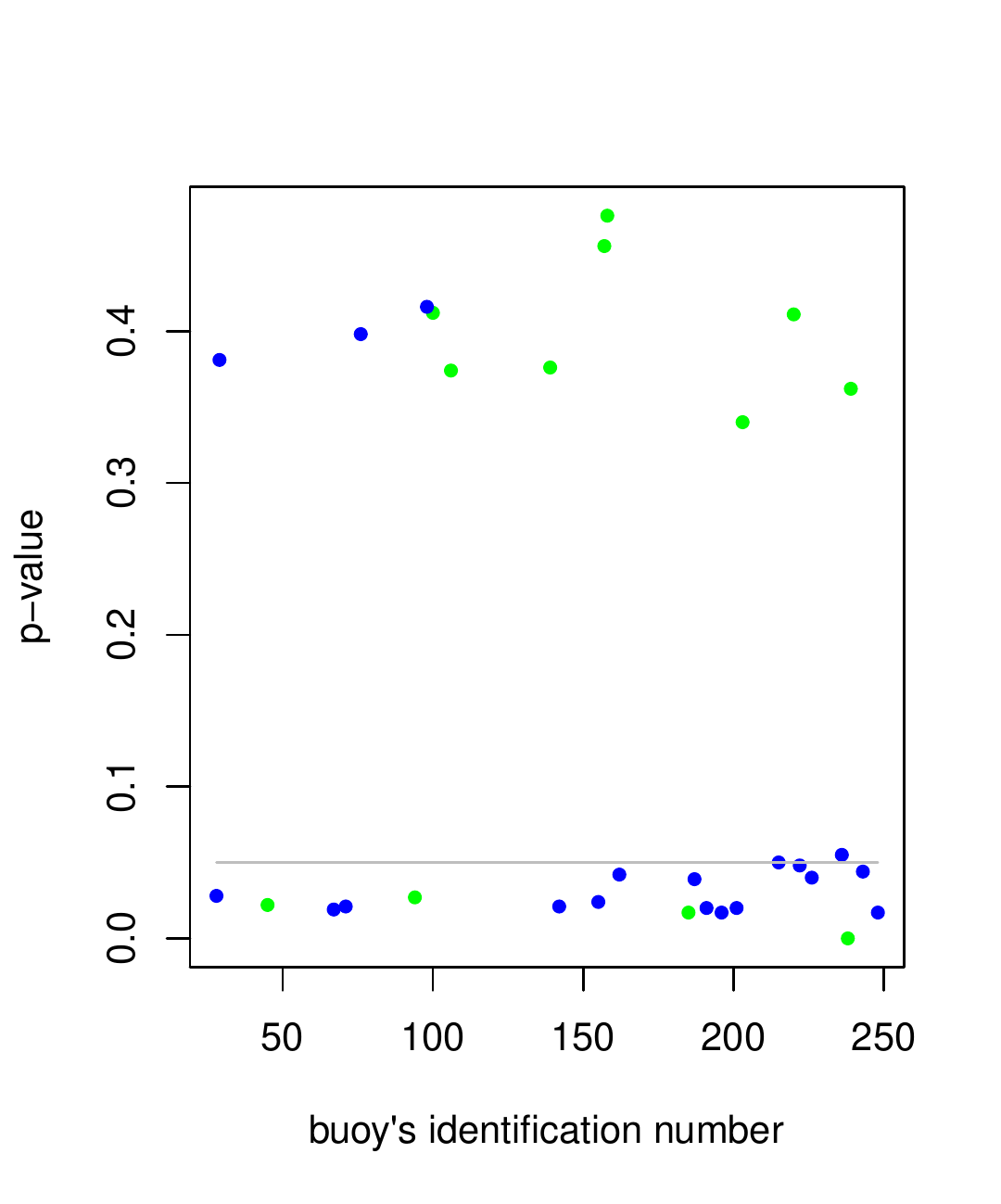}
   \end{center}
  \caption{Left panel: FDR corresponding to the studied buoys  in Table  \ref{tab:tab3}.
  Right panel: P-value corresponding to the buoys studied in Table \ref{tab:tab4}. P-values obtained with the random projection test based on the one-dimensional Lovato and Velasco test are represented in color  blue and those based on the Epps test are in color green.  The line $y=.05$ is displayed in both panels in  color grey.
  }\label{Plot_tab56}
\end{figure}

To better illustrate the findings, the results of Table \ref{tab:tab3} are summarized in the left plot of Figure \ref{Plot_tab56}. The $x$-axis represents the buoy's identification number while the $y$-axis displays the obtained FDR for dependent tests. A grey line at $y=0.05$ is drawn to show what buoys have a FDR above or below that value, which result in a rejection of the null hypothesis \eqref{03}. It can observe that there are three FDR that are just above 0.05. They correspond   to buoys  028, 071 and 185.

 \begin{table}[t]
\caption{P-values (column 2) resulting of applying the random projection test for each buoy (column 1) with a FDR adjusted p-value larger than 0.05 in Table \ref{tab:tab3}. The parameters (columns 3 and 4) and the one-dimensional test (column 5) used in performing the random projection test are included for each buoy. 
}
 \begin{center}
   \setlength{\tabcolsep}{8mm}
\begin{tabular}{rrrrrrrr}
  \hline
 \bf buoy&  \bf p-value &  \bf param1 &  \bf param2  &  \bf  test  \\ \hline \\
  248 &  \bf .017   & 100  &    1  & L.-V. 
\\
  243 &  \bf .044   & 100   &   1  & L.-V.
\\
  239 &   .362    &  100    &  1  & Epps
\\
  238 &  \bf .000    &100   &   1  & Epps
\\
  236 &   .055   &   2& 7  & L.-V.
\\
  226 &  \bf .040   & 100  &    1  & L.-V.
\\
  222 &  \bf  .048 &  2 &7  & L.-V.
\\
  220 &   .411&	2& 7  & Epps
\\
  215 &  \bf .050   & 100   & 1  & L.-V.
\\
 203 &   .340	&	2& 7  & Epps
\\
 201 &  \bf .020	&100  &    1  & L.-V.
\\
 196 &   \bf.017  &    2  &    7  & L.-V.
\\
 191 &  \bf .020&	100  &    1  & L.-V.
\\
 187 & \bf  .039	& 2   &   7  & L.-V.
\\
 185 & \bf  .017  &  100  &    1  & Epps
\\
 162 & \bf  .042  &  100   &   1  & L.-V.
\\
 158 &   .476&	2 &7  & Epps
\\
 157 &   .456	&2 &7  & Epps
\\
 155 & \bf  .024&	100   &   1  & L.-V.
\\
 142 & \bf  .021	&100   &   1  & L.-V.
\\
 139 &   .376&	2 &7  & Epps
\\
 106 &   .374&	2 &7  & Epps
\\
 100 &   .412&	100   &  1  & Epps
\\
 098 &   .416	&100    &  1  & Epps
\\
 094 &  \bf .027	&100   &   1  & L.-V.
\\
 076 &   .398	&100   &   1  & Epps
\\
 071 &   \bf.021  &  100   &   1  & L.-V.
\\
 067 & \bf .019	&100     & 1  & L.-V.
\\
 045 &  \bf .022	&100    &  1  & L.-V.
\\
 029 &   .381	&100    &  1  & Epps
\\
 028 &   \bf.028   & 100    &  1  & L.-V.
   \\
\hline
\end{tabular}
\end{center}
\label{tab:tab4} 
\end{table}

In what follows it is pursued a further study in the 31 buoys for which there  is yet no evidence to reject the null hypothesis of Gaussianity, displayed in \eqref{04}. This further study consists in applying the random projection test. In applying it, the information in Table  \ref{tab:tab3} obtained when applying de Epps and Lobato and Velasco tests is used. For instance, if 
for one of these two tests the associated p-value is smaller than 0.05, it is made use of that test and the parameters $(100,1)$ in computing the random projection test. Remember that, as commented in Section \ref{metho}, making use of the $(100,1)$ parameters results in a projected time series similar to the original one. 

The results of applying the random projection test are reported in Table \ref{tab:tab4}. There it can be observed that the random projection test is able to reject the null hypothesis of Gaussianity in 19 out of the 31 buoys, which results in a total of 50 rejections out of 62 (the 80.65$\%$). The p-values that result in a rejection are highlighted in bold. The p-value associated to buoy 215 has been highlighted because it takes value 4.966946e-02, which in the table has been rounded to 0.050.
The table also provides  the parameters used to compute the projection and the test applied to it. 

The results in Table \ref{tab:tab4} have been summarized in the right plot of Figure \ref{Plot_tab56}. There,  the p-values larger and smaller than 0.05 can be clearly observed;  and that there is a p-value just above 0.05, the one corresponding to buoy 236. The p-values where the Lovato and Velasco test is used in performing the random projection test are colored in blue. Those in which the random projection test makes use of the Epps test are colored in green.

\bibliographystyle{imsart-nameyear}	
\bibliography{Olas}

\begin{thebibliography}{45}

\bibitem[\protect\citeauthoryear{Azaïs, León and Ortega}{2005}]{Azais2005}
\begin{barticle}[author]
\bauthor{\bsnm{Azaïs},~\bfnm{Jean-Marc}\binits{J.-M.}},
  \bauthor{\bsnm{León},~\bfnm{José~R.}\binits{J.~R.}} \AND
  \bauthor{\bsnm{Ortega},~\bfnm{Joaquín}\binits{J.}}
(\byear{2005}).
\btitle{Geometrical characteristics of Gaussian sea waves}.
\bjournal{Journal of Applied Probability}
\bvolume{42}
\bpages{407–425}.
\bdoi{10.1239/jap/1118777179}
\end{barticle}
\endbibitem

\bibitem[\protect\citeauthoryear{Battjes and Groenendijk}{2000}]{Battjes}
\begin{barticle}[author]
\bauthor{\bsnm{Battjes},~\bfnm{Jurjen~A}\binits{J.~A.}} \AND
  \bauthor{\bsnm{Groenendijk},~\bfnm{Heiko~W}\binits{H.~W.}}
(\byear{2000}).
\btitle{Wave height distributions on shallow foreshores}.
\bjournal{Coastal Engineering}
\bvolume{40}
\bpages{161-182}.
\bdoi{https://doi.org/10.1016/S0378-3839(00)00007-7}
\end{barticle}
\endbibitem

\bibitem[\protect\citeauthoryear{Benetazzo et~al.}{2015}]{Benetazzo2015}
\begin{barticle}[author]
\bauthor{\bsnm{Benetazzo},~\bfnm{A.}\binits{A.}},
  \bauthor{\bsnm{Barbariol},~\bfnm{F.}\binits{F.}},
  \bauthor{\bsnm{Bergamasco},~\bfnm{F.}\binits{F.}},
  \bauthor{\bsnm{Torsello},~\bfnm{A.}\binits{A.}},
  \bauthor{\bsnm{Carniel},~\bfnm{S.}\binits{S.}} \AND
  \bauthor{\bsnm{Sclavo},~\bfnm{M}\binits{M.}}
(\byear{2015}).
\btitle{Observation of Extreme Sea Waves in a Space–Time Ensemble}.
\bjournal{Journal of Physical Oceanography}
\bvolume{45}
\bpages{2261-2275}.
\end{barticle}
\endbibitem

\bibitem[\protect\citeauthoryear{Benjamini and Hochberg}{1995}]{Benjamin1995}
\begin{barticle}[author]
\bauthor{\bsnm{Benjamini},~\bfnm{Yoav}\binits{Y.}} \AND
  \bauthor{\bsnm{Hochberg},~\bfnm{Yosef}\binits{Y.}}
(\byear{1995}).
\btitle{Controlling the False Discovery Rate: A Practical and Powerful Approach
  to Multiple Testing}.
\bjournal{Journal of the Royal Statistical Society. Series B (Methodological)}
\bvolume{57}
\bpages{289--300}.
\end{barticle}
\endbibitem

\bibitem[\protect\citeauthoryear{Benjamini and Yekutieli}{2001}]{Benjamin2001}
\begin{barticle}[author]
\bauthor{\bsnm{Benjamini},~\bfnm{Yoav}\binits{Y.}} \AND
  \bauthor{\bsnm{Yekutieli},~\bfnm{Daniel}\binits{D.}}
(\byear{2001}).
\btitle{The Control of the False Discovery Rate in Multiple Testing under
  Dependency}.
\bjournal{The Annals of Statistics}
\bvolume{29}
\bpages{1165--1188}.
\end{barticle}
\endbibitem

\bibitem[\protect\citeauthoryear{Boccotti}{1989}]{Boccotti}
\begin{barticle}[author]
\bauthor{\bsnm{Boccotti},~\bfnm{P.}\binits{P.}}
(\byear{1989}).
\btitle{On Mechanics of Irregular Gravity Waves}.
\bjournal{Atti della Accademia Nazionale dei Lincei, Memorie}
\bvolume{19}
\bpages{110-170}.
\end{barticle}
\endbibitem

\bibitem[\protect\citeauthoryear{Box and Pierce}{1970}]{Box}
\begin{barticle}[author]
\bauthor{\bsnm{Box},~\bfnm{G.~E.~P.}\binits{G.~E.~P.}} \AND
  \bauthor{\bsnm{Pierce},~\bfnm{David~A.}\binits{D.~A.}}
(\byear{1970}).
\btitle{Distribution of Residual Autocorrelations in Autoregressive-Integrated
  Moving Average Time Series Models}.
\bjournal{Journal of the American Statistical Association}
\bvolume{65}
\bpages{1509-1526}.
\bdoi{10.1080/01621459.1970.10481180}
\end{barticle}
\endbibitem

\bibitem[\protect\citeauthoryear{Casas‐Prat and
  Holthuijsen}{2010}]{CasasPrat}
\begin{barticle}[author]
\bauthor{\bsnm{Casas‐Prat},~\bfnm{M.}\binits{M.}} \AND
  \bauthor{\bsnm{Holthuijsen},~\bfnm{L.~H.}\binits{L.~H.}}
(\byear{2010}).
\btitle{Short‐term statistics of waves observed in deep water}.
\bjournal{Journal of Geophysical Research: Oceans}
\bvolume{115 (C9)}.
\bdoi{https://doi.org/10.1029/2009JC005742}
\end{barticle}
\endbibitem

\bibitem[\protect\citeauthoryear{Coleman}{1974}]{Coleman1974}
\begin{binbook}[author]
\bauthor{\bsnm{Coleman},~\bfnm{Rodney}\binits{R.}}
(\byear{1974}).
\btitle{What is a Stochastic Process?}
In \bbooktitle{Stochastic Processes}
\bpages{1--5}.
\bpublisher{Springer Netherlands}, \baddress{Dordrecht}.
\bdoi{10.1007/978-94-010-9796-3_1}
\end{binbook}
\endbibitem

\bibitem[\protect\citeauthoryear{Collins et~al.}{2014}]{collins2014recording}
\begin{barticle}[author]
\bauthor{\bsnm{Collins},~\bfnm{Clarence~O}\binits{C.~O.}},
  \bauthor{\bsnm{Lund},~\bfnm{Bj{\"o}rn}\binits{B.}},
  \bauthor{\bsnm{Waseda},~\bfnm{Takuji}\binits{T.}} \AND
  \bauthor{\bsnm{Graber},~\bfnm{Hans~C}\binits{H.~C.}}
(\byear{2014}).
\btitle{On recording sea surface elevation with accelerometer buoys: lessons
  from ITOP (2010)}.
\bjournal{Ocean dynamics}
\bvolume{64}
\bpages{895--904}.
\end{barticle}
\endbibitem

\bibitem[\protect\citeauthoryear{Cuesta-Albertos et~al.}{2007}]{Cuesta2007}
\begin{barticle}[author]
\bauthor{\bsnm{Cuesta-Albertos},~\bfnm{J.~A.}\binits{J.~A.}},
  \bauthor{\bparticle{del} \bsnm{Barrio},~\bfnm{E.}\binits{E.}},
  \bauthor{\bsnm{Fraiman},~\bfnm{R.}\binits{R.}} \AND
  \bauthor{\bsnm{Matrán},~\bfnm{C.}\binits{C.}}
(\byear{2007}).
\btitle{The Random Projection Method in Goodness of Fit for Functional Data}.
\bjournal{Computational Statistics \& Data Analysis}
\bvolume{51}
\bpages{4814 - 4831}.
\bdoi{https://doi.org/10.1016/j.csda.2006.09.007}
\end{barticle}
\endbibitem

\bibitem[\protect\citeauthoryear{D'Agostino and Stephens}{1986}]{Dagostino1987}
\begin{barticle}[author]
\bauthor{\bsnm{D'Agostino},~\bfnm{Ralph~B.}\binits{R.~B.}} \AND
  \bauthor{\bsnm{Stephens},~\bfnm{Michael~A.}\binits{M.~A.}}
(\byear{1986}).
\btitle{Goodness-of-fit techniques}.
\bjournal{Quality and Reliability Engineering International}
\bvolume{3}
\bpages{71-71}.
\bdoi{10.1002/qre.4680030121}
\end{barticle}
\endbibitem

\bibitem[\protect\citeauthoryear{Epps}{1987}]{epps1987}
\begin{barticle}[author]
\bauthor{\bsnm{Epps},~\bfnm{T.~W.}\binits{T.~W.}}
(\byear{1987}).
\btitle{Testing That a Stationary Time Series is Gaussian}.
\bjournal{The Annals of Statistics}
\bvolume{15}
\bpages{1683--1698}.
\bdoi{10.1214/aos/1176350618}
\end{barticle}
\endbibitem

\bibitem[\protect\citeauthoryear{Forristall}{1978}]{Forristall1978}
\begin{barticle}[author]
\bauthor{\bsnm{Forristall},~\bfnm{G.~Z.}\binits{G.~Z.}}
(\byear{1978}).
\btitle{On the statistical distribution of wave heights in a storm}.
\bjournal{Journal of Geophysical Research: Oceans}
\bvolume{83}
\bpages{2353-2358}.
\end{barticle}
\endbibitem

\bibitem[\protect\citeauthoryear{Haver}{1987}]{Haver1987}
\begin{barticle}[author]
\bauthor{\bsnm{Haver},~\bfnm{Sverre}\binits{S.}}
(\byear{1987}).
\btitle{On the joint distribution of heights and periods of sea waves}.
\bjournal{Ocean Engineering}
\bvolume{14}
\bpages{359-376}.
\bdoi{https://doi.org/10.1016/0029-8018(87)90050-3}
\end{barticle}
\endbibitem

\bibitem[\protect\citeauthoryear{Hessner, Reichert and
  Hutt}{2007}]{hessner2007sea}
\begin{binproceedings}[author]
\bauthor{\bsnm{Hessner},~\bfnm{Katrin}\binits{K.}},
  \bauthor{\bsnm{Reichert},~\bfnm{Konstanze}\binits{K.}} \AND
  \bauthor{\bsnm{Hutt},~\bfnm{Belmont-Lower}\binits{B.-L.}}
(\byear{2007}).
\btitle{Sea surface elevation maps obtained with a nautical X-Band
  radar--Examples from WaMoS II stations}.
In \bbooktitle{10th International Workshop on Wave Hindcasting and Forecasting
  and Coastal Hazard Symposium, North Shore, Oahu, Hawaii}
\bpages{11--16}.
\end{binproceedings}
\endbibitem

\bibitem[\protect\citeauthoryear{Hokimoto and Shimizu}{2014}]{hokimoto2014non}
\begin{barticle}[author]
\bauthor{\bsnm{Hokimoto},~\bfnm{Tsukasa}\binits{T.}} \AND
  \bauthor{\bsnm{Shimizu},~\bfnm{Kunio}\binits{K.}}
(\byear{2014}).
\btitle{A non-homogeneous hidden Markov model for predicting the distribution
  of sea surface elevation}.
\bjournal{Journal of applied statistics}
\bvolume{41}
\bpages{294--319}.
\end{barticle}
\endbibitem

\bibitem[\protect\citeauthoryear{J.}{2006}]{Pitman}
\begin{bincollection}[author]
\bauthor{\bsnm{J.},~\bfnm{Pitman}\binits{P.}}
(\byear{2006}).
\btitle{Combinatorial Stochastic Processes}.
In \bbooktitle{Lectures from the 32nd Summer School on Probability Theory held
  in Saint-Flour}
\bpublisher{Springer}.
\end{bincollection}
\endbibitem

\bibitem[\protect\citeauthoryear{Jishad, Yadhunath and
  Seelam}{2021}]{Jishad2021}
\begin{barticle}[author]
\bauthor{\bsnm{Jishad},~\bfnm{M.}\binits{M.}},
  \bauthor{\bsnm{Yadhunath},~\bfnm{E.~M.}\binits{E.~M.}} \AND
  \bauthor{\bsnm{Seelam},~\bfnm{Jaya~Kumar}\binits{J.~K.}}
(\byear{2021}).
\btitle{Wave height distribution in unsaturated surf zones}.
\bjournal{Regional Studies in Marine Science}
\bvolume{44}
\bpages{101708}.
\bdoi{https://doi.org/10.1016/j.rsma.2021.101708}
\end{barticle}
\endbibitem

\bibitem[\protect\citeauthoryear{Karmpadakis, Swan and
  Christou}{2020}]{Karmpadakis2020}
\begin{barticle}[author]
\bauthor{\bsnm{Karmpadakis},~\bfnm{I.}\binits{I.}},
  \bauthor{\bsnm{Swan},~\bfnm{C.}\binits{C.}} \AND
  \bauthor{\bsnm{Christou},~\bfnm{M.}\binits{M.}}
(\byear{2020}).
\btitle{Assessment of wave height distributions using an extensive field
  database}.
\bjournal{Coastal Engineering}
\bvolume{157}
\bpages{103630}.
\bdoi{https://doi.org/10.1016/j.coastaleng.2019.103630}
\end{barticle}
\endbibitem

\bibitem[\protect\citeauthoryear{Klopman}{1996}]{Klopman}
\begin{barticle}[author]
\bauthor{\bsnm{Klopman},~\bfnm{G.}\binits{G.}}
(\byear{1996}).
\btitle{Extreme wave heights in shallow water}.
\bjournal{WL|Delft Hydraulics}
\bvolume{Report H2486}.
\end{barticle}
\endbibitem

\bibitem[\protect\citeauthoryear{Kozachenko et~al.}{2016}]{KOZACHENKO201671}
\begin{bincollection}[author]
\bauthor{\bsnm{Kozachenko},~\bfnm{Yuriy}\binits{Y.}},
  \bauthor{\bsnm{Pogorilyak},~\bfnm{Oleksandr}\binits{O.}},
  \bauthor{\bsnm{Rozora},~\bfnm{Iryna}\binits{I.}} \AND
  \bauthor{\bsnm{Tegza},~\bfnm{Antonina}\binits{A.}}
(\byear{2016}).
\btitle{2 - Simulation of Stochastic Processes Presented in the Form of
  Series}.
In \bbooktitle{Simulation of Stochastic Processes with Given Accuracy and
  Reliability}
(\beditor{\bfnm{Yuriy}\binits{Y.}~\bsnm{Kozachenko}},
  \beditor{\bfnm{Oleksandr}\binits{O.}~\bsnm{Pogorilyak}},
  \beditor{\bfnm{Iryna}\binits{I.}~\bsnm{Rozora}} \AND
  \beditor{\bfnm{Antonina}\binits{A.}~\bsnm{Tegza}}, eds.)
\bpages{71-104}.
\bpublisher{Elsevier}.
\bdoi{https://doi.org/10.1016/B978-1-78548-217-5.50002-7}
\end{bincollection}
\endbibitem

\bibitem[\protect\citeauthoryear{Kwiatkowski et~al.}{1992}]{KppsI1992}
\begin{barticle}[author]
\bauthor{\bsnm{Kwiatkowski},~\bfnm{Denis}\binits{D.}},
  \bauthor{\bsnm{Phillips},~\bfnm{Peter C.~B.}\binits{P.~C.~B.}},
  \bauthor{\bsnm{Schmidt},~\bfnm{Peter}\binits{P.}} \AND
  \bauthor{\bsnm{Shin},~\bfnm{Yongcheol}\binits{Y.}}
(\byear{1992}).
\btitle{Testing the Null Hypothesis of Stationarity Against the Alternative of
  a Unit Root: How sure Are We that Economic Time Series Have a Unit Root?}
\bjournal{Journal of Econometrics}
\bvolume{54}
\bpages{159 - 178}.
\bdoi{https://doi.org/10.1016/0304-4076(92)90104-Y}
\end{barticle}
\endbibitem

\bibitem[\protect\citeauthoryear{Lobato and Velasco}{2004}]{Lobato2004}
\begin{barticle}[author]
\bauthor{\bsnm{Lobato},~\bfnm{Ignacio}\binits{I.}} \AND
  \bauthor{\bsnm{Velasco},~\bfnm{Carlos}\binits{C.}}
(\byear{2004}).
\btitle{A simple Test of Normality for Time Series}.
\bjournal{Econometric Theory}
\bvolume{20}
\bpages{671-689}.
\bdoi{10.1017/S0266466604204030}
\end{barticle}
\endbibitem

\bibitem[\protect\citeauthoryear{Longuet-Higgins}{1980}]{Longuet1980}
\begin{barticle}[author]
\bauthor{\bsnm{Longuet-Higgins},~\bfnm{Michael~S.}\binits{M.~S.}}
(\byear{1980}).
\btitle{On the distribution of the heights of sea waves: Some effects of
  nonlinearity and finite band width}.
\bjournal{Journal of Geophysical Research: Oceans}
\bvolume{85}
\bpages{1519-1523}.
\bdoi{https://doi.org/10.1029/JC085iC03p01519}
\end{barticle}
\endbibitem

\bibitem[\protect\citeauthoryear{Mendes and Scotti}{2021}]{Mendes2021}
\begin{barticle}[author]
\bauthor{\bsnm{Mendes},~\bfnm{S.}\binits{S.}} \AND
  \bauthor{\bsnm{Scotti},~\bfnm{A.}\binits{A.}}
(\byear{2021}).
\btitle{The Rayleigh-Haring-Tayfun distribution of wave heights in deep water}.
\bjournal{Applied Ocean Research}
\bvolume{113}
\bpages{102739}.
\bdoi{https://doi.org/10.1016/j.apor.2021.102739}
\end{barticle}
\endbibitem

\bibitem[\protect\citeauthoryear{Mendez, Losada and Medina}{2004}]{Mendez}
\begin{barticle}[author]
\bauthor{\bsnm{Mendez},~\bfnm{Fernando~J}\binits{F.~J.}},
  \bauthor{\bsnm{Losada},~\bfnm{Inigo~J}\binits{I.~J.}} \AND
  \bauthor{\bsnm{Medina},~\bfnm{Raul}\binits{R.}}
(\byear{2004}).
\btitle{Transformation model of wave height distribution on planar beaches}.
\bjournal{Coastal Engineering}
\bvolume{50}
\bpages{97-115}.
\bdoi{https://doi.org/10.1016/j.coastaleng.2003.09.005}
\end{barticle}
\endbibitem

\bibitem[\protect\citeauthoryear{Mori, Liu and Yasuda}{2002}]{Mori}
\begin{barticle}[author]
\bauthor{\bsnm{Mori},~\bfnm{Nobuhito}\binits{N.}},
  \bauthor{\bsnm{Liu},~\bfnm{Paul~C}\binits{P.~C.}} \AND
  \bauthor{\bsnm{Yasuda},~\bfnm{Takashi}\binits{T.}}
(\byear{2002}).
\btitle{Analysis of freak wave measurements in the Sea of Japan}.
\bjournal{Ocean Engineering}
\bvolume{29}
\bpages{1399-1414}.
\bdoi{https://doi.org/10.1016/S0029-8018(01)00073-7}
\end{barticle}
\endbibitem

\bibitem[\protect\citeauthoryear{Muraleedharan
  et~al.}{2007}]{Muraleedharan2007}
\begin{barticle}[author]
\bauthor{\bsnm{Muraleedharan},~\bfnm{G.}\binits{G.}},
  \bauthor{\bsnm{Rao},~\bfnm{A.~D.}\binits{A.~D.}},
  \bauthor{\bsnm{Kurup},~\bfnm{P.~G.}\binits{P.~G.}},
  \bauthor{\bsnm{Nair},~\bfnm{N.~Unnikrishnan}\binits{N.~U.}} \AND
  \bauthor{\bsnm{Sinha},~\bfnm{Mourani}\binits{M.}}
(\byear{2007}).
\btitle{Modified Weibull distribution for maximum and significant wave height
  simulation and prediction}.
\bjournal{Coastal Engineering}
\bvolume{54}
\bpages{630-638}.
\end{barticle}
\endbibitem

\bibitem[\protect\citeauthoryear{Naess}{1985}]{Naess}
\begin{barticle}[author]
\bauthor{\bsnm{Naess},~\bfnm{Arvid}\binits{A.}}
(\byear{1985}).
\btitle{On the distribution of crest to trough wave heights}.
\bjournal{Ocean Engineering}
\bvolume{12}
\bpages{221-234}.
\bdoi{https://doi.org/10.1016/0029-8018(85)90014-9}
\end{barticle}
\endbibitem

\bibitem[\protect\citeauthoryear{Nieto-Reyes}{2021}]{Olas1}
\begin{barticle}[author]
\bauthor{\bsnm{Nieto-Reyes},~\bfnm{Alicia}\binits{A.}}
(\byear{2021}).
\btitle{On the Non-Gaussianity of the Height of Sea Waves}.
\bjournal{Journal of Marine Science and Engineering}
\bvolume{9}.
\bdoi{10.3390/jmse9121446}
\end{barticle}
\endbibitem

\bibitem[\protect\citeauthoryear{Nieto-Reyes, Cuesta-Albertos and
  Gamboa}{2014}]{nietoreyes2014}
\begin{barticle}[author]
\bauthor{\bsnm{Nieto-Reyes},~\bfnm{Alicia}\binits{A.}},
  \bauthor{\bsnm{Cuesta-Albertos},~\bfnm{Juan~Antonio}\binits{J.~A.}} \AND
  \bauthor{\bsnm{Gamboa},~\bfnm{Fabrice}\binits{F.}}
(\byear{2014}).
\btitle{A Random-Projection Based Test of Gaussianity for Stationary
  Processes}.
\bjournal{Computational Statistics \& Data Analysis}
\bvolume{75}
\bpages{124 - 141}.
\bdoi{https://doi.org/10.1016/j.csda.2014.01.013}
\end{barticle}
\endbibitem

\bibitem[\protect\citeauthoryear{Pena-Sanchez, M{\'e}rigaud and
  Ringwood}{2018}]{pena2018short}
\begin{barticle}[author]
\bauthor{\bsnm{Pena-Sanchez},~\bfnm{Yerai}\binits{Y.}},
  \bauthor{\bsnm{M{\'e}rigaud},~\bfnm{Alexis}\binits{A.}} \AND
  \bauthor{\bsnm{Ringwood},~\bfnm{John~V}\binits{J.~V.}}
(\byear{2018}).
\btitle{Short-term forecasting of sea surface elevation for wave energy
  applications: The autoregressive model revisited}.
\bjournal{IEEE Journal of Oceanic Engineering}
\bvolume{45}
\bpages{462--471}.
\end{barticle}
\endbibitem

\bibitem[\protect\citeauthoryear{Perron}{1988}]{Perron1988}
\begin{barticle}[author]
\bauthor{\bsnm{Perron},~\bfnm{Pierre}\binits{P.}}
(\byear{1988}).
\btitle{Trends and Random Walks in Macroeconomic Time Series: Further Evidence
  From a New Approach}.
\bjournal{Journal of Economic Dynamics and Control}
\bvolume{12}
\bpages{297 - 332}.
\bdoi{https://doi.org/10.1016/0165-1889(88)90043-7}
\end{barticle}
\endbibitem

\bibitem[\protect\citeauthoryear{Reichert, Dannenberg and van~den
  Boom}{2010}]{reichert2010x}
\begin{binproceedings}[author]
\bauthor{\bsnm{Reichert},~\bfnm{Konstanze}\binits{K.}},
  \bauthor{\bsnm{Dannenberg},~\bfnm{Jens}\binits{J.}} \AND
  \bauthor{\bparticle{van~den} \bsnm{Boom},~\bfnm{Henk}\binits{H.}}
(\byear{2010}).
\btitle{X-Band radar derived sea surface elevation maps as input to ship motion
  forecasting}.
In \bbooktitle{OCEANS'10 IEEE SYDNEY}
\bpages{1--7}.
\bpublisher{IEEE}.
\end{binproceedings}
\endbibitem

\bibitem[\protect\citeauthoryear{Rozanov}{1967}]{zbMATH03244325}
\begin{bmisc}[author]
\bauthor{\bsnm{Rozanov},~\bfnm{Y.~A.}\binits{Y.~A.}}
(\byear{1967}).
\btitle{Stationary random processes}.
\bhowpublished{San {Francisco}-{Cambridge}-{London}-{Amsterdam}: {Holden}-{Day}
  1967. 211 p. (1967).}
\end{bmisc}
\endbibitem

\bibitem[\protect\citeauthoryear{Said and Dickey}{1984}]{dickey1984}
\begin{barticle}[author]
\bauthor{\bsnm{Said},~\bfnm{Said~E.}\binits{S.~E.}} \AND
  \bauthor{\bsnm{Dickey},~\bfnm{David~A.}\binits{D.~A.}}
(\byear{1984}).
\btitle{Testing for Unit Roots in Autoregressive-Moving Average Models of
  Unknown Order}.
\bjournal{Biometrika}
\bvolume{71}
\bpages{599-607}.
\bdoi{10.1093/biomet/71.3.599}
\end{barticle}
\endbibitem

\bibitem[\protect\citeauthoryear{Schulz-Stellenfleth and
  Lehner}{2004}]{schulz2004measurement}
\begin{barticle}[author]
\bauthor{\bsnm{Schulz-Stellenfleth},~\bfnm{Johannes}\binits{J.}} \AND
  \bauthor{\bsnm{Lehner},~\bfnm{Susanne}\binits{S.}}
(\byear{2004}).
\btitle{Measurement of 2-D sea surface elevation fields using complex synthetic
  aperture radar data}.
\bjournal{IEEE Transactions on Geoscience and Remote Sensing}
\bvolume{42}
\bpages{1149--1160}.
\end{barticle}
\endbibitem

\bibitem[\protect\citeauthoryear{Srokosz}{1986}]{srokosz1986joint}
\begin{barticle}[author]
\bauthor{\bsnm{Srokosz},~\bfnm{MA}\binits{M.}}
(\byear{1986}).
\btitle{On the joint distribution of surface elevation and slopes for a
  nonlinear random sea, with an application to radar altimetry}.
\bjournal{Journal of Geophysical Research: Oceans}
\bvolume{91}
\bpages{995--1006}.
\end{barticle}
\endbibitem

\bibitem[\protect\citeauthoryear{Srokosz and
  Longuet-Higgins}{1986}]{srokosz1986skewness}
\begin{barticle}[author]
\bauthor{\bsnm{Srokosz},~\bfnm{MA}\binits{M.}} \AND
  \bauthor{\bsnm{Longuet-Higgins},~\bfnm{MS}\binits{M.}}
(\byear{1986}).
\btitle{On the skewness of sea-surface elevation}.
\bjournal{Journal of Fluid Mechanics}
\bvolume{164}
\bpages{487--497}.
\end{barticle}
\endbibitem

\bibitem[\protect\citeauthoryear{Stansell}{2004}]{Stansell4}
\begin{barticle}[author]
\bauthor{\bsnm{Stansell},~\bfnm{Paul}\binits{P.}}
(\byear{2004}).
\btitle{Distributions of freak wave heights measured in the North Sea}.
\bjournal{Applied Ocean Research}
\bvolume{26}
\bpages{35-48}.
\bdoi{https://doi.org/10.1016/j.apor.2004.01.004}
\end{barticle}
\endbibitem

\bibitem[\protect\citeauthoryear{Stansell}{2005}]{Stansell5}
\begin{barticle}[author]
\bauthor{\bsnm{Stansell},~\bfnm{Paul}\binits{P.}}
(\byear{2005}).
\btitle{Distributions of extreme wave, crest and trough heights measured in the
  North Sea}.
\bjournal{Ocean Engineering}
\bvolume{32}
\bpages{1015-1036}.
\bdoi{https://doi.org/10.1016/j.oceaneng.2004.10.016}
\end{barticle}
\endbibitem

\bibitem[\protect\citeauthoryear{Tayfun}{1990}]{Tayfun}
\begin{barticle}[author]
\bauthor{\bsnm{Tayfun},~\bfnm{M.~Aziz}\binits{M.~A.}}
(\byear{1990}).
\btitle{Distribution of Large Wave Heights}.
\bjournal{Journal of Waterway, Port, Coastal, and Ocean Engineering}
\bvolume{116}
\bpages{686-707}.
\bdoi{10.1061/(ASCE)0733-950X(1990)116:6(686)}
\end{barticle}
\endbibitem

\bibitem[\protect\citeauthoryear{van Vledder}{1991}]{Vledder}
\begin{barticle}[author]
\bauthor{\bparticle{van} \bsnm{Vledder},~\bfnm{G.~P.}\binits{G.~P.}}
(\byear{1991}).
\btitle{Modification of the Glukhovskiy distribution}.
\bjournal{Technical report, Delft Hydraulics}
\bvolume{Report H1203}.
\end{barticle}
\endbibitem

\bibitem[\protect\citeauthoryear{Wu et~al.}{2016}]{Wu}
\begin{barticle}[author]
\bauthor{\bsnm{Wu},~\bfnm{Yanyun}\binits{Y.}},
  \bauthor{\bsnm{Randell},~\bfnm{David}\binits{D.}},
  \bauthor{\bsnm{Christou},~\bfnm{Marios}\binits{M.}},
  \bauthor{\bsnm{Ewans},~\bfnm{Kevin}\binits{K.}} \AND
  \bauthor{\bsnm{Jonathan},~\bfnm{Philip}\binits{P.}}
(\byear{2016}).
\btitle{On the distribution of wave height in shallow water}.
\bjournal{Coastal Engineering}
\bvolume{111}
\bpages{39-49}.
\bdoi{https://doi.org/10.1016/j.coastaleng.2016.01.015}
\end{barticle}
\endbibitem

\end{thebibliography}

\end{document}